\documentclass[letter,scriptaddress,twocolumn, prl,showkeys]{revtex4}
	\usepackage{amsmath}
	\usepackage{makeidx}
	\usepackage{amsfonts}
	\usepackage[ansinew]{inputenc}
	\usepackage[usenames,dvipsnames]{pstricks}
	\usepackage{epsfig}
	\usepackage{pst-grad} 
	\usepackage{pst-plot} 
	\usepackage[colorlinks,hyperindex]{hyperref}
	\hypersetup
	{
		colorlinks,%
		citecolor=black,%
		linkcolor=black,%
		urlcolor=black,%
	}

\makeindex

\begin{document}

\title { Microscopic Insights into Dynamical Heterogeneity in a Lane-forming Liquid}

\author{Suman Dutta}
\email{Email:sumand@bose.res.in}
\affiliation{Department of Chemical Biological and Macro-molecular Sciences\\
S. N. Bose National Centre for Basic Sciences \\ Block-JD, Sector-III, Salt Lake, Kolkata 700106 \\ India.\\
}

\date{12 November, 2017 }

\begin{abstract}
Dynamical heterogeneity (DH) in non-equilibrium systems is a topic of profound interest yet an
open question. In a model system of constantly driven oppositely charged binary colloidal suspension, we explore DH in a model lane-forming system using BD simulations. We show that the system undergoes
 structural and dynamical cross-over using spatio-temporal correlation functions. For small field,
 the structural relaxation is homogeneous while it is heterogeneous for
 sufficiently high field. In order to explore the heterogeneity, we track and tag the particles to compute
 partial structures that relax at different rates, in which heterogeneity has its maximum in the intermediate state.

\end{abstract}

\maketitle

 Driven away from equilibrium, soft materials exhibit fascinating phenomena, ranging from dynamic patterns to active self-assembly, and often generate technological applications in diverse areas of science \cite{soft1}. Yet the underlying microscopic description of the driven systems is mostly unexplored, for the particle dynamics depends explicitly on the structural changes, and external field competes with inter-particle interactions\cite{soft2}. This interplay between structure and dynamics governs the transport processes and hence develops widespread interests in a variety of inter-disciplinary subjects bridging physics, chemistry, biology and engineering\cite{soft1,soft2,rev1}. 

Laning \cite{rev1,rev2,lane1,lane2,lane3,lane3a,netz, lane4,lane5,lane6,lane7,lane8,lane9, lane9a,lane10,lane11,lane12} is one such example realized in a host of simple systems like army ants\cite{ant}, pedestrian movements\cite{ped}, granular media\cite{granule}, dusty plasma\cite{plasma}, dipolar microswimmers \cite{swimmers} and is often considered as a generic model of non-equilibrium systems where two species of particles are driven against each other \cite{ped,granule,ant,plasma,lane1,lane2,lane3,lane3a,netz, lane4,lane5,lane6,lane7,lane8,lane9,lane10,lane11,lane12}. 
Colloids mimic the phenomenon: applying a constant electric field, the system of binary charges crosses over from a homogeneous mixture to a state with columnar lanes of likely charged particles elongated parallel to the field\cite{soft1,soft2,rev1,rev2,lane1,lane2,lane3,lane3a, netz, lane4,lane5,lane6,lane7,lane8,lane9,lane9a,lane10,lane11,lane12}. This laning transition is accompanied by enhanced diffusion in the plane perpendicular to the applied field when particles are surrounded by those of the opposite charges which have been further quantified in geometric terms\cite{lane11}. There is a `locked-in' situation as the system gets more heterogeneous in the lane state due to the coexistence of both slow and fast mobile particles\cite{lane8}. This leads to a dynamical heterogeneity (DH) \cite{lane8} and heterogeneity in structural relaxation\cite{lane10} as the primitive lanes grow with field \cite{lane9}. The correlations decay exponentially in the transverse field and algebraically in the direction of the field\cite{lane9a,lane10,lane12}. Interestingly, laning occurs via the {\it pre-lane} state with anomalous dynamical responses due to a heterogeneity in diffusion\cite{lane10}, like that in super-cooled liquids\cite{smk}, yet the nature of dynamics differs at a single particle level. Although predicted as a generic feature in soft materials, DH and heterogeneous structural relaxation lack macroscopic realization and farther microscopic insights\cite{nat_mat}. Thus, exploring dynamical and structural heterogeneity in the system would not only contribute in developing technological applications but would also aid inputs into the theoretical understanding of collective response in a class of non-equilibrium systems with similar inherent structural heterogeneity being influenced by field.

In this paper, we consider a simple system of a binary mixture of oppositely charged colloids in presence of a constant electric field. We show that with increasing field, there is an increase in the effective attraction between particles of same charge and the effective repulsion between opposite charges, as the lanes grow monotonically with the strength of the field. However, the spread in the cluster distribution behaves non-monotonically on the field strength. The Overlap function\cite{cd} $Q(t)$, captures the slowing down in the transverse plane as the strength of the field increases. We observe a crossover in dynamics: from an initial faster relaxation in the homogeneous state to the lane state accompanied by the transverse plane slowing down via the `pre-lane' state with heterogeneous relaxation. The Dynamic susceptibility\cite{smk-cd} $\chi_{4}(t)$, crosses over from a unimodal form in the homogeneous state to a form with double peaks in the lane state while the pre-lane state show broadening in $\chi_{4}(t)$. From particle tagged description of the density autocorrelations, we show that these heterogeneous relaxations, indeed, manifest novel heterogeneous structures that have a spectrum of timescales of relaxation in which the heterogeneity has its maximum, again, in the intermediate state. 

We take an equi-molar binary mixture of positively ($N_{+}$) and negatively ($N_{-}$) charged colloidal particles of diameter $\sigma(=1\mu m$)($N_{+}$=$N_{-}$=2000) in a cubic box of length ($L=21.599\sigma $) dissolved in a medium with viscosity($\eta=0.1cP$) at temperature $T(=298K)$ as in Ref. \cite{lane10}. The pair-interaction between two particles located at positions $\vec{r_{i}}$ and $\vec{r_{j}}$, with charges $q_{i}$ and $q_{j}$, respectively is given by $V(r_{ij} )= V_{SC} + V_{Repulsion}$ with  $ V_{SC} = V_{0} [q_{i} q_{j} /(1+\kappa \sigma /2)^{2} ][\exp (-\kappa \sigma ((r_{ij} /\sigma )-1))/r_{ij} /\sigma ] $ and $V_{repulsion} =\varepsilon [(\sigma /r_{ij} )^{12} -(\sigma /r_{ij} )^{6} ]+\frac{1}{4}$ for $r_{ij} <2^{1/6} \sigma $  and zero, elsewhere \cite{lane5} with $r_{ij} =|\mathop{r_{i}}\limits^{\to } -\mathop{r_{j}}\limits^{\to } |$. Here, $\kappa $ is the inverse screening length, $V_{0 }$ the interaction strength parameter and $\varepsilon =4\left|q\right|^{2} V_{0} (1+\kappa \sigma /2)^{2} $ with $q_{i}=q_{j}=q$ \cite{lane5}. We use $\kappa \sigma (=5.0)$and $V_{0}^{*} =\left|q\right|^{2} V_{0} /k_{B} T(=50.0)$. The BD simulations \cite{erm} are carried out using discretized form of Langevin's equation in over-damped limit with integration time step $\Delta t=0.00005\tau_{\beta}$. We use $\Gamma (=3\pi \sigma \eta )$, the viscous damping and $\mathop{F_{i}}\limits^{\to} (t)$ the fluctuating force with variance $2D_{0} \delta _{\alpha \beta } \delta _{ij} \delta (t-t')$ where $\alpha$, $\beta $ denote the cartesian components and $D_{0}$ the Einstein-Stokes Diffusion coefficient with $\Gamma D_{0} =k_{B} T$, $k_{B} $ being the Boltzmann constant. The simulations use $\tau _{\beta } =(\sigma^{2} /D_{0} $ ) as unit time, $\sigma$ as the length unit and $k_{B} T$ as the energy unit. We switch on the electric field $f(=\left|q\right|f_{0} \sigma /k_{B} T)$ once we equilibrate the system with $f=0$ from random configurations for $50\tau_{\beta}$. The field is kept on for $100\tau_{\beta}$ so that for all $f$ (within the observation window), the system reaches steady state where statistics are stored for $50\tau_{\beta}$. We generate $N_{T}(=20)$ set of trajectories with different initial configurations. During our analysis we average over these $N_{T}$ Brownian trajectories generated using different seeds with different initially equilibrated configurations. We set field strengths in accordance to the study in Ref. \cite{lane10} where the chosen three values of $f$ correspond to three different dynamical states namely homogeneous state, lane state and the `pre-lane' state, the intermediate {\it anomalous} state \cite{lane10}. 

  The projection of the lanes in the transverse plane to the field are shown in Fig.1 via the particle configurations. Fig. 1(a) show a homogeneous mixture of opposite charges for $f=50$. The plots are very similar to that for $f=0$, the equilibrium configurations. However, for $f=150$, we observe tiny domains of likely charged particles [Fig. 1(b)]. On increasing $f$ further, these domains takes the form of network like structures [Fig. 1(c)] proliferated in z direction as lanes as in previous studies \cite{lane5,lane10}. The laning tendency in the system is monitored by the lane order parameter, $\Phi$, defined in Ref.\cite{lane5}. In steady states, $\Phi$ shows a steady value $\Phi_{S}$, with fluctuations. With increasing $f$, $\Phi_{S}$ grows as the laning tendency in the system increases. Though there is a steady flow, the lanes continue to re-order in the transverse plane as seen in different steady state patterns (not shown) at a particular $f$. This is affirmed via the complicated decay in the autocorrelation of $\Phi$ [Fig. 5 in Appendix].

\begin{figure}[h]
\includegraphics[angle=0,scale=0.09]{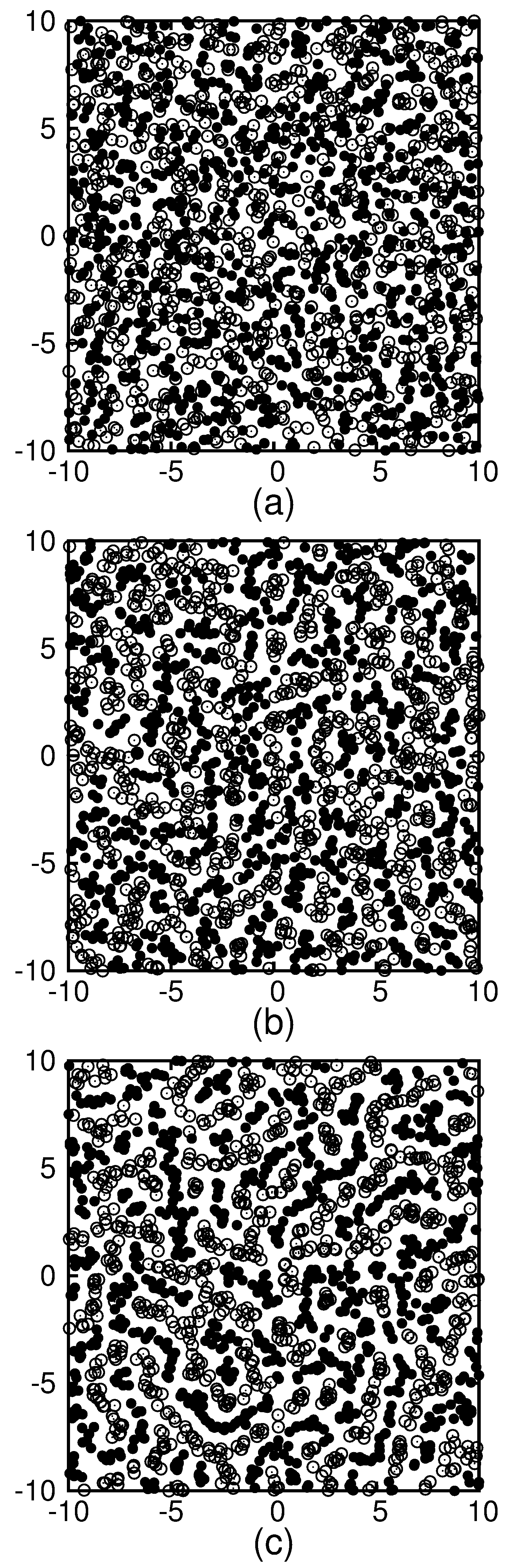}
\caption{Projection of the particles in plane transverse to the field for (a)$f=50$ (b)$f=150$ (c)$f=300$. Here open circles and filled circles denote $+ve$ and $-ve$ charges respectively.}
\end{figure}

The steady state structural rearrangements are governed by effective interactions mediated by the interacting lanes present in the system. The pair distribution functions \cite{allen,ef_int} for two particles at a separation $r_{\bot}$ in the plane transverse to the applied field are given by $g^{(++)}(r_{\bot})$ and $g^{(+-)}(r_{\bot})$ for the like and cross species pairs respectively. We observe $g^{(++)}(r_{\bot})\sim g^{(--)}(r_{\bot})$ and $g^{(+-)}(r_{\bot})\sim g^{(-+)}(r_{\bot})$ (Data not shown). In the transverse plane, the Effective interaction ($V_{eff}^{(++)}(r_{\bot})$) between a pair of $+ve$ particles in presence of other particles is given by the relation: $g^{(++)}(r_{\bot})\sim \exp(-\beta V_{eff}^{(++)}(r_{\bot}))$. This leads to $V_{eff}^{(++)}(r_{\bot})\sim -\beta \ln g^{(++)}(r_{\bot})$ \cite{ef_int}. Similarly, we have $V_{eff}^{(+-)}(r_{\bot})\sim -\beta \ln g^{(+-)}(r_{\bot})$. In Fig. 2(a) we show the dependence of $V_{eff}^{(+-)}(r_{\bot})$ (Main Panel) and $V_{eff}^{(++)}(r_{\bot})$ (Inset) on $r_{\bot}$. For $f\neq0$, we observe a peak in $V_{eff}^{(+-)}(r_{\bot})$ and a dip in $V_{eff}^{(++)}(r_{\bot})$ for $r_{\bot}\approx 0$ that grows with increasing $f$. This indicates that with increasing $f$, the system experiences an enhanced effective attraction between like charge-pairs while an increased effective repulsion between oppositely charged pairs. The increase in both the effective interactions follow the monotonic growth in steady-state structural order parameter with increasing $f$, reported in earlier studies \cite{lane10,lane5}. Here, $V_{eff}^{(++)}(r_{\bot})$ deviates from $V^{(++)}(r_{\bot})$ for $f\ne 0$ due to increase in collective behavior of a particular species. This excess contribution is due to the increased many body contribution arising out of the competing particle interaction coupled to the applied field. Hence, for a high $f$, effective interaction between two like charges is more influenced by the neighboring like particles in the same lane while the same for two opposite charges is dominated by the effective interaction of two lanes of the opposite charges. Since $V_{eff}^{(++)}(r_{\bot})\neq V^{(++)}(r_{\bot})$ and $V_{eff}^{(+-)}(r_{\bot})\neq V^{(+-)}(r_{\bot})$ for $f\neq 0$, $V_{eff}^{(++)}(r_{\bot}) \sim V^{(++)}(r_{\bot})+ \delta V^{(++)} (r_{\bot})$ where $\delta V^{(++)} (r_{\bot})$ contains contributions due to direct correlation \cite{ef_int} in particle pairs which, here, is coupled to the applied $f$. For $f=0$, one ends up with $\delta V^{(++)} (r_{\bot})\approx \delta V^{(+-)} (r_{\bot}) \approx 0$ in the low density limit. 

\begin{figure}[h]
\includegraphics[angle=0,scale=0.18]{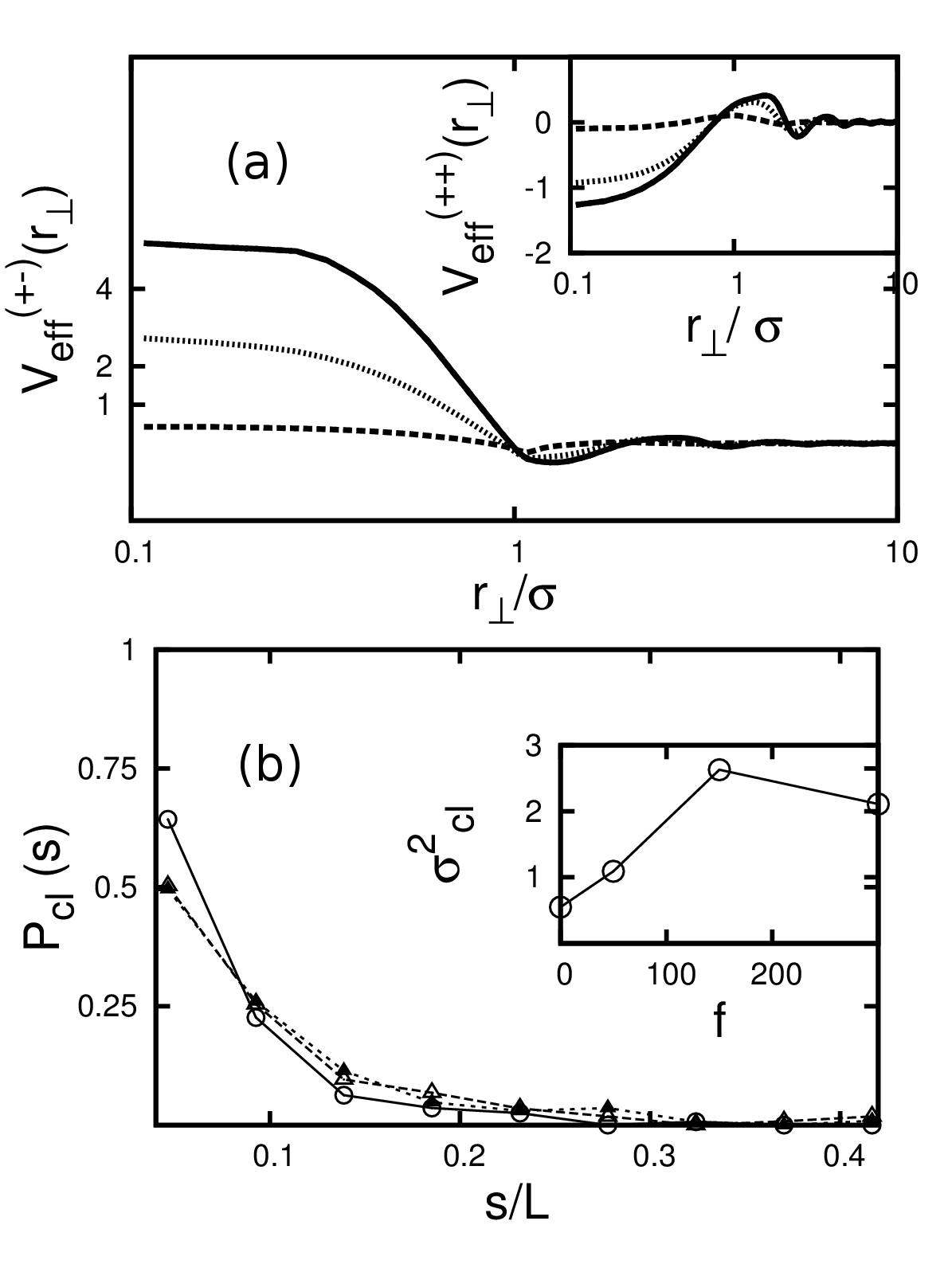}
\caption{(a) Effective Interactions: $V_{eff}^{(+-)}(r_{\bot})$ vs $r_{\bot}$ for $f=$ 50(dashed line),150 (dotted line)and 300 (bold line) Inset. $V_{eff}^{(++)}(r_{\bot})$ vs $r_{\bot}$ for $f=$ 50(dashed line),150 (dotted line) and 300 (bold line)(b) Dependence of $P_{cl}(s)$ on $s/L$ for $f=$50(open circles with solid line), 150(filled triangles and dot-dashed line) ,300(open triangles and dotted line). Lines are only guide to eyes. Inset. $\sigma^{2}_{cl}$ vs $f$. }
\end{figure}

Motivated by Ref.\cite{lane3a}, in order to find the associated size distribution of the lanes present in the system, we compute the probability of a particle to be a part of a particular cluster of size $s$. In a cluster of likely charged particles, we subsequently add particles of the same species within a critical separation \cite{allen} in three dimensions, $r_{cl}^{(++)}$ (for +ve particles) and $r_{cl}^{(--)}$ (for -ve particles) to obtain the size of the cluster $s$. This is repeated for all the particles of the same charge to obtain the cluster size distribution. The probability distribution, $P_{cl}(s)$ is computed for a particular value of $r_{cl}^{(\pm \pm)}(=1.4\sigma)$. We plot $P_{cl}(s)$ for the $+ve$ charges for $r_{cl}^{(\pm \pm)}(=1.4\sigma)$ as a function of $s$ for different $f$ in Fig. 2(b). For $f=0$, $P_{cl}(s)$ shows high peak at $s=1$ indicating isolated clusters. The situation remains somewhat similar for $f=50$ when the system is mostly spanned by small clusters. However, for both $f=150$ and $f=300$, the initial peak in $P_{cl}(s)$ decrease significantly while the probability increase for higher $s$. Mean cluster size of the system, $<s_{cl}>$ is given by $\int sP_{cl}(s)ds$ while average number of attached neighbors  $<\xi _{cl}> \sim \int (s-1)P_{cl}(s-1)ds$ excluding the particle in the reference. The spread in the distribution is given by $\sigma_{cl}^{2}=<\xi^{2}_{cl}>-<\xi _{cl}>^{2}$. In Inset Fig.2(b), with increasing $f$, $\sigma_{cl}^{2}$ show a maximum at $f=150$. The trend of $\sigma_{cl}^{2}$ is similar to the trend of heterogeneity in diffusion in the same system\cite{lane10}. 

So we attempt to relate the structural changes with the changes in the dynamic properties. To understand the underlying structural rearrangements, we look for the dynamical relaxation of these structures. Such structural relaxation is generally interpreted via the self Overlaps\cite{smk-cd}, given by 
\begin{equation}
\tilde{q}^{(+)}(t)\sim \frac{1}{N_{+}} \sum_{i=1} ^{N_{+}} \psi(|\vec{r}(t_{0})-\vec{r}(t+t_{0})|)
\end{equation}
Here, $\psi(r)=1$ if $r\leq 0.3\sigma$ and $\psi(r)=0$ elsewhere. To compute such overlaps between particle configurations in the orthogonal plane, separated by time $t$, we evaluate $\tilde{q}^{(\pm)}_{(\bot)}(t)$ for both $+ve$ and $-ve$ charges respectively. Both charges behave identically, so we focus on $\tilde{q}^{(+)}_{(\bot)}(t)$. We compute $\tilde{q}^{(+)}(t)$ for several steady state configurations separated by $t$. The probability distribution function of $\tilde{q}^{(+)}(t)$ is given by $P(\tilde{q}^{(+)}(t))$. We observe $P (\tilde{q}_{(\bot)}^{(+)}(0)) \sim \delta(\tilde{q}_{(\bot)}^{(+)}(0) - N_{+})$. For $t \neq 0$, the peaks in $P (\tilde{q}_{(\bot)}^{(+)}(t))$ shift with increaseing $t$. In Fig. 3(a) we show the dependence of $P (\tilde{q}_{(\bot)}^{(+)}(t))$ for various $f$ for a typical $t=0.05\tau_{\beta}$. Form small $t$, $P (\tilde{q}_{(\bot)}^{(+)}(t))$ is a Gaussian. For $f=50$, at $t=0.05$, $P (\tilde{q}_{(\bot)}^{(+)}(t))$ has a peak at $\tilde{q}_{(\bot)}^{(+)}(t)\approx 0.89$.  This peak shifts to lower values of $\tilde{q}_{(\bot)}^{(+)}(t)$ ($\approx 0.87$) for $f=150$ and then again shifts to the higher values of $\tilde{q}_{(\bot)}^{(+)}(t)$($\approx 0.91$) for $f=300$ indicating a non-monotonic trend. This behaviour may be associated with the dynamical changes in the system reported in Ref. \cite{lane10}. 

The shift of peaks in $P(\tilde{q}_{(\bot)}^{(+)}(t))$ is given by the Overlap Function\cite{smk-cd}, 

\begin{equation}
Q_{(\bot)}^{(+)}(t) \sim \int \tilde{q}_{(\bot)}^{(+)}(t)P(\tilde{q}_{(\bot)}^{(+)}(t))d \tilde{q}_{(\bot)}^{(+)}(t)).
\end{equation}

 For all $f$, $Q^{(+)}_{(\bot)}(0)=1$  and they decay monotonically with $t$ [Fig. 3(b)]. However, the decay rates in $Q^{(+)}_{(\bot)}(t)$ vary with the strength of $f$. $f=50$ shows a relatively fast decay in $Q^{(+)}_{(\bot)}(t)$. The rate decrease with $f$ indicating a slowing down in the transverse plane as in Ref. \cite{netz}. For $f=300$, we observe that $Q^{(+)}_{(\bot)}(t)\approx 0.25$ which is twice the value $(\approx 0.12)$ of that for $f=50$ at $t=1\tau_{\beta}$. We find an intermediate value $\approx 0.18$ in $Q^{(+)}_{(\bot)}$ for $f=150$. Also there exists a crossover in $Q^{(+)}_{(\bot)}$. For $f=50$, $Q^{(+)}_{(\bot)}(t)\sim t^{\alpha}$ where $\alpha \approx -0.86$ while we find  $Q^{(+)}_{(\bot)}(t)\sim e^{-t^{- \beta}}$ in the time window $10\tau_{\beta}<t<40\tau_{\beta}$ for $f=150$ with $\beta\approx 0.37$[Inset. Fig. 6 in Appendix]. Beyond this, $Q^{(+)}_{(\bot)}(t)$ shows a power law dependence, like in aggregating liquids \cite{td} with $\alpha\approx -0.52$ for $f=300$ where the fitting time window is limited to $1\tau_{\beta}<t<20\tau_{\beta}$ (Main panel Fig. 6 in Appendix).

\begin{figure}[h]
\includegraphics[angle=0,scale=0.12]{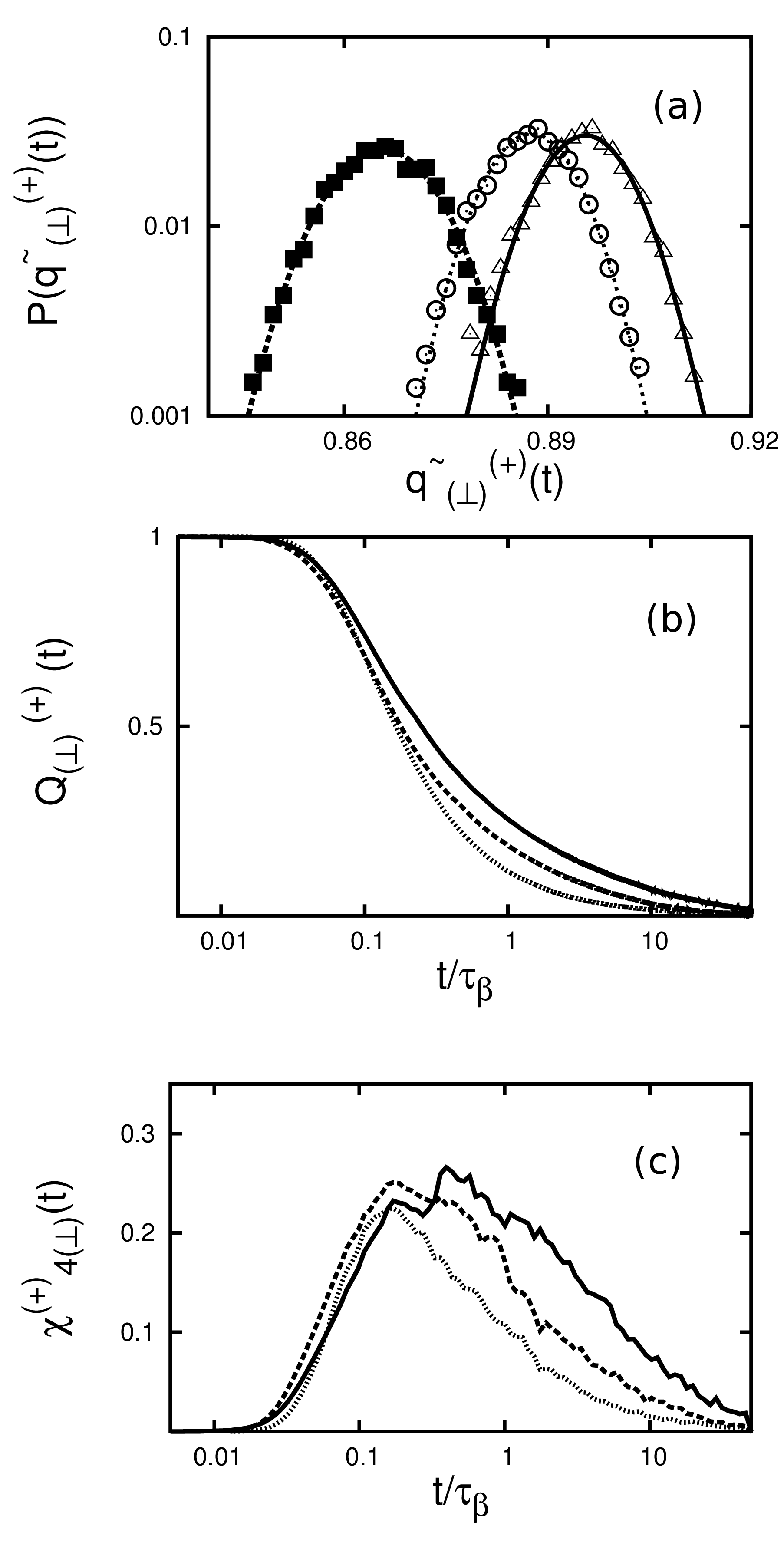}
\caption{(a) Distribution of Overlaps, $P(\tilde{q}_{(\bot)}^{(+)}(t))$ for different $\tilde{q}_{(\bot)}^{(+)}(t)$ is shown for $t=0.05 \tau_{\beta}$: $f=50$ (open circles), $f=150$ (filled squares) and $f=300$(open triangles). Lines show the fitted Gaussian curves. (b) Dependences of $Q_{(\bot)}^{(+)}(t)$ on $t$ for three regimes: fast-segregation $f=50$ (dotted line), mixed relaxation $f=150$ (dashed line) and slow relaxation $f=300$ (solid line) (c) Dynamical Susceptibility, $\chi^{(+)}_{4}(t)$ for three regimes: $f=$50 (dotted line),150 (dashed line) and 300 (solid line)}
\end{figure}

Structural heterogeneity induces slowing down of dynamics in a system \cite{td}. In order to understand the underlying structural response we compute the dynamical susceptibility, $\chi^{(\pm)}_{4}(t)$ which is given in terms of the dynamic fluctuations in $Q(t)$ \cite{cd,smk-cd}, 
\begin{equation}
\chi^{(\pm)}_{4}(t)=<Q^{(\pm)2}(t)>-<Q^{(\pm)}(t)>^{2}.
\end{equation}
It peaks at $t=\tau_{4}$ which is proportional to the structural relaxation time, $\tau$ \cite{smk-cd}. We show the evolution of $\chi^{(+)}_{4(\bot)}(t)$ with $t$ for different $f$ in Fig. 3(c). $\chi^{(+)}_{4(\bot)}(t)$ for $f=0$ grows with $t$ and shows a peak at $t=\tau_{4}$ (data not shown) as in normal liquid\cite{smk-cd}. For $f=50$, the peak shifts to lower value of $t$ than that for $f=0$ indicating initial phase segregation due to the faster relaxation. For $f=150$, $\chi^{(+)}_{4(\bot)}(t)$ grows and broadens with no prominent peak, showing the coexisting time-scales of structural relaxation with comparable magnitudes in the system. On increasing $f$ further, $\chi^{(+)}_{4(\bot)}(t)$ shows two distinct peaks for $f=300$ depicting heterogeneity. 

 In the homogeneous state, the dynamics is entirely governed by the fast particles driven by the field. The peak in $\chi^{(+)}_{4(\bot)}(t)$ for $f=50$ corresponds to relaxation by the faster particles in the system. In contrast, the slow dynamics in the lane state is associated with the particles in the proliferated lanes. This results in the predominant peak in $\chi^{(+)}_{4(\bot)}(t)$ at higher $t$ for $f=300$ while the peak at lower $t$ is entirely due to the fast particles. The dynamics in the intermediate state experiences a competition between the two. Hence, the broadening in $\chi^{(+)}_{4(\bot)}(t)$ for $f=150$. This is consistent with the data of heterogeneity in structural relaxation in the {\it pre-lane} state\cite{lane10}. 

Now we investigate whether it is possible to link this heterogeneity with the heterogeneity density relaxation reported recently\cite{lane10}. In this process, we compute the probability distribution of square of the particle displacements, $P(\Delta r^{2}_{\bot}, t)$, in the plane transverse to the applied field in a given time interval, $t$. We observe $P^{(+)}(\Delta r^{2}_{\bot}, t)\approx P^{(-)}(\Delta r^{2}_{\bot}, t)(=P(\Delta r^{2}_{\bot}, t))$. For $t=0$, $P(\Delta r^{2}_{\bot}, t)$ has a peak at $\Delta r^{2}_{\bot}=0$. We plot $P(\Delta r^{2}_{\bot}, t)$ with $\Delta r^{2}_{\bot}$ for $f=150$(Main Panel) and $f=300$(Inset) in Fig. 4(a). For $t > 0$, the the peaks in $P(\Delta r^{2}_{\bot}, t)$ locates at $\Delta r^{2}_{\bot}=\Delta R_{P}$ with value $\Delta P_{v}$. With increasing $t$, we observe that the peak at $\Delta R_{P}$ shifts to the higher values of $\Delta r^{2}_{\bot}$ while $\Delta P_{v}$ decays due to diffusion  [see Fig. 7 in Appendix]. The decay rates depend on the strength of $f$. This trend is similar to the decay of $Q_{(\bot)}^{(+)}(t)$ and the distinct van Hove functions as in Ref. \cite{lane10}, affirming the increase in transverse plane slowing down in the system. Now, in order to identify the fast and slow relaxing particles within a particular species, we tag particles as "slow relaxing" particle (S) if it has square of the displacement $\Delta r^{2}_{\bot}(t )<\Delta R_{P}(t)$ in the time window $t$. Similarly we tag the particles as "fast relaxing" (F) if $\Delta r^{2}_{\bot}(t) \geq \Delta R_{P}(t)$. Thus we count the no F and S particles ($N^{(+)}_{F}(t)$ and $N^{(+)}_{S}(t)$ respectively) of $+ve$ and $-ve$ charges respectively in a given time interval $t$.

Since the system consists of S and F particles of $+ve$ and $-ve$ charges, there exist various possibilites structure between two particles can relax. For example, among $+ve$ charged particles, the way a S particle relaxes in the vicinity of other S particles is different to the same in the vicinity of F particles. Thus, there exists six such possibilities. We now analyze the density relaxation of these structures via  the distinct van Hove function which infers how a particle relax in the vicinity of other particles in the system. The density relaxation of a positively charged S particle in the vicinity of other S particles of the same species in the time interval $t$, is given by \cite{ef_int}

\begin{align}
G_{D(S,S)}^{(++)} &(r,t)=2/(N^{(+)}_{S}(t)(N^{(+)}_{S}(t)-1))\nonumber \\
&<\sum _{i,j=1}^{N^{(+)}_{S}(t)}\sum _{j\neq i} \delta (r +|\vec{R_{j}}(t)-\vec{R_{i}} (0)| > 
\end{align}

 Similarly, that between two such S particles of opposite charge is given by \cite{ef_int}
 
\begin{align}
&G_{D(S, S)}^{(+-)}(r,t)=1/(N^{(+)}_{S}(t)(N^{(-)}_{S}(t))\nonumber \\
&<\sum _{i=1}^{N^{(+)}_{S}(t)}\sum _{j=1}^{N^{(-)}_{S}(t)} \delta (r +|\vec{R_{j}}(t)-\vec{R_{i}} (0)| > 
\end{align}

The other possibilities are $G_{D(F,F)}^{(+\pm)} (r,t)$ and $G_{D(F,S)}^{(+\pm)} (r,t)$. Since the lanes interact among themselves in the transverse plane we compute these quantities in the plane transverse to the applied field to revisit and explore the heterogeneity in structural relaxation reported earlier in Ref. \cite{lane10}.   

\begin{figure}[h]
\includegraphics[angle=0,scale=0.12]{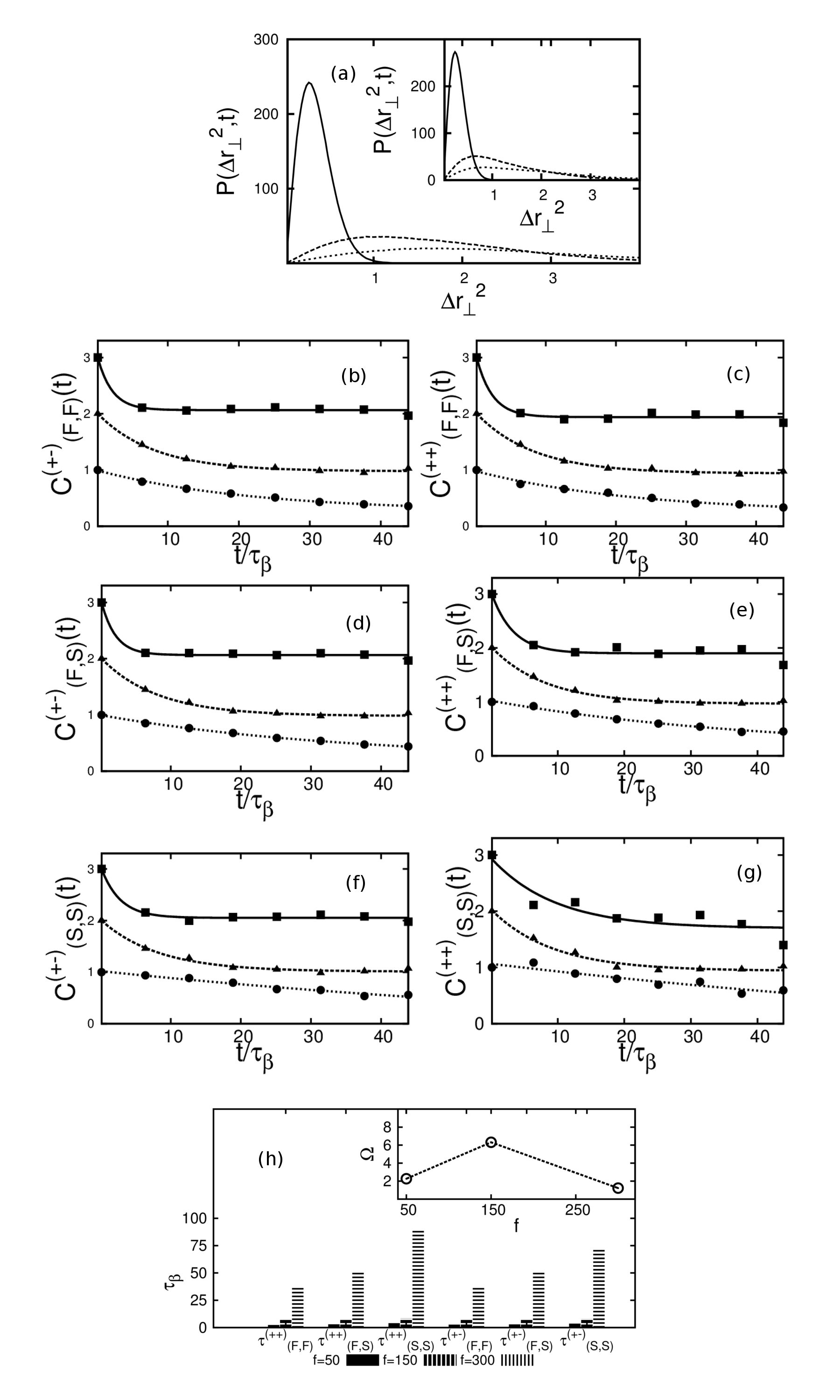}
\caption{ (a) $P(\Delta r^{2}_{\bot}, t)$ vs $\Delta r^{2}_{\bot}$ for $t=0.25\tau_{\beta}$(solid line), $12.5\tau_{\beta}$ (dashed line)and $=25\tau_{\beta}$(dotted line) for $f=150$ (Main Panel) and $f=300$ (Inset)(b)$C_{(F,F)}^{(+-)} (t)$, (c)$C_{(F,F)}^{(++)} (t)$,(d) $C_{(F,S)}^{(+-)} (t)$, (e)$C_{(F,S)}^{(++)} (t)$ (f) $C_{(S,S)}^{(+-)} (t)$  (g)$C_{(S,S)}^{(++)} (t)$ as a function of $t$ for $f=50$(filled squares, shown with vertical offset 2), $f=150$(filled triangles, shown with vertical offset 1) and $f=300$ (filled circles). Solid, dashed and dotted lines show fitted lines for $f=50$(shown with vertical offset 2), $f=150$ (shown with vertical offset 1) and $f=300$ respectively.(h)$\tau_{(F,F)}^{(++)}$, $\tau_{(F,S)}^{(++)}$, $\tau_{(S,S)}^{(++)}$, $\tau_{(F,F)}^{(+-)}$, $\tau_{(F,S)}^{(+-)}$ and $\tau_{(S,S)}^{(+-)}$  for $f=50$, $f=150$ and $f=300$. Inset: Dependence of $\Omega$ on $f$}
\end{figure}

To obtain the structural relaxation at wave-vector $q_{\bot}(=2\pi/L)$, we compute the Fourier transforms of these quantities are given by $G_{D (S,S)}^{(+\pm)} (q_{\bot},t)$, $G_{D(F,F)}^{(+\pm)} (q_{\bot},t)$ and $G_{D(F,S)}^{(+\pm)} (q_{\bot},t)$ respectively. The decay profiles of the peaks in $G_{D}^{(++)} (q_{\bot},t)$ and dips in $G_{D}^{(+-)} (q_{\bot},t)$ at wave vector $q_{0}$ is given as 

\begin{equation}
C_{(M,N)}^{(+\pm)} (t)= \frac{\pm G_{D (M,N)}^{(+\pm)} (q_{0},t)\mp 1}{\pm G_{D(M,N)}^{(+\pm)} (q_{0},0)\mp 1}
\end{equation}

as in Ref. \cite{lane10}. Here $M$ and $N$ are indices that stand for both $S$ and $F$.

 We show evolution of $C_{(M,N)}^{(+\pm)} (t)$ with $t$ for various $f$ in Fig. 5(b-g). For all $f$, we find $C_{(S,S)}^{(+\pm)}(0)=C_{(F,S)}^{(+\pm)}(0)=C_{(F,F)}^{(+\pm)}(0)=1$. For all $M$ and $N$, $C_{(M,N)}^{(+\pm)} (t)$ monotonically decrease with $t$. However, the decay rates depends upon the combination of $M$ and $N$ at a particular $f$. For $f=50$, $C_{(F,F)}^{(++)} (t)\approx C_{(F,F)}^{(+-)} (t)\approx C_{(F,S)}^{(++)} (t)$ for small $t$ and both show fast relaxation in the observation time window. However, $C_{(F,S)}^{(+-)} (t)$ slows down for high $t$, while $C_{(S,S)}^{(++)} (t)$ show distinct changes. For $f=150$,  both $C_{(F,F)}^{(+\pm)} (t)$ and  $C_{(F,S)}^{(+\pm)} (t)$ show fast decay while  $C_{(S,S)}^{(+\pm)} (t)$ show relatively slower decay. The scenario changes in case of $f=300$. Both $C_{(S,S)}^{(++)} (t)$ and $C_{(S,S)}^{(+-)} (t)$ show slow decay while both $C_{(F,F)}^{(++)} (t)$ and $C_{(F,F)}^{(+-)} (t)$ show fast decay. Moreover, we observe $C_{(F,S)}^{(++)} (t)\approx C_{(F,S)}^{(+-)} (t)$ for $f=300$.  We find $C_{(M,N)}^{(+\pm)} (t) \sim \exp(-t/\tau^{(+\pm)}_{(M,N)})$ (fits are shown in Fig.4(b-g) ) where the the timescales of relaxation are given by $\tau^{(+\pm)}_{(M,N)}$ and are shown in the main panel of Fig. 4(h). For $f=50$, we find $\tau^{(++)}_{(F,F)} < \tau^{(+-)}_{(F,F)} < \tau^{(+-)}_{(F,S)} < \tau^{(++)}_{(F,S)} < \tau^{(+-)}_{(S,S)} < \tau^{(++)}_{(S,S)}$. The trend changes for $f=150$ as $\tau^{(++)}_{(F,S)} \approx \tau^{(+-)}_{(F,F)}$ and $\tau^{(++)}_{(F,S)} < \tau^{(+-)}_{(F,S)}$. However, these timescales are relatively close in magnitudes. However, for $f=300$, we observe $\tau^{(++)}_{(F,F)} < \tau^{(+-)}_{(F,F)} < \tau^{(++)}_{(F,S)} < \tau^{(+-)}_{(F,S)} < \tau^{(+-)}_{(S,S)} < \tau^{(++)}_{(S,S)}$. Here, $\tau^{(+\pm)}_{(S,S)}$ increase significantly with increasing $f$. The measure of heterogeneity is given by $\Omega=\frac{1}{(\sigma_{\tau}/\mu_{\tau})}$ with $\sigma_{\tau}^{2}=\sum_{M,N} (\tau_{M,N} - \mu_{\tau})^{2}$ and $\mu_{\tau}=1/6 \sum_{M,N} \tau_{M,N}$, the mean relaxation time. $\Omega$ show non-monotonic dependence on $f$ [Inset Fig. 5(h)], like similar non-monotonic behaviour in $\sigma_{cl}^{2}$. This is also quite similar to the behavior of heterogeneity in diffusion in the same system\cite{lane10}. 

 Reports \cite{lane1,lane8} show that the laning transition is accompanied by an initial increase in the diffusion. The increasing drift enforces the particles to move through lanes. But the particle motion in the transverse direction decrease as the bigger lanes starts to appear. This is due to the increasing effective attraction between the like charges and repulsion between the cross charges. The particles in the same lane cause the slow dynamics while the cross-lane movements are associated with faster diffusion. In the homogeneous state, the pre-dominant attraction between the opposite charges results into the faster diffusion while in the lane state, the inter-lane movements become low probable leading to a slowing down. The state for intermediate strength of the field experiences a competition between the two movements. Hence, we observe the onset of multiple time-scales in the system. This lead to DH and the associated anomalous dynamical responses in the {\it pre-lane} state\cite{lane10}. The dynamics in lane state is thus mostly governed by the slow particles in the same lane while a low populating fast particles are associated with the fast diffusion tail in the diffusion spectrum in Ref. \cite{lane10}. Thus, we observe the separation of time-scales in the lane state via the double peaks in $\chi_{4}(t)$. The double peaks in $\chi_{4}(t)$ has been previously seen in super-cooled liquids and it has been linked with short time $\beta$-relaxation \cite{short-beta} where the time-scales of the slow and fast particles are widely separated.

In conclusion, in a driven mixture of oppositely charged colloid, we probe simultaneous relaxations of the lanes as the system approaches the laning transition from an initial homogeneous mixture. With increasing field, the lanes proliferate and their cumulative interactions grow monotonically with increasing field. However, the cluster size distribution evolves non-monotonically. From two and four point time correlation functions, we observe heterogeneity in structural relaxations. Since the individual lanes relax differently and simultaneously, the system shows a heterogeneous structural relaxation with competing timescales in the intermediate state while in the lane state, there is a separation of time-scales in distinct peaks due to increased proportion of slow particles. But unlike glasses, these time scales are not separated by orders of magnitude although both show slowing down of dynamics. The intermediate heterogeneous state involves mixed relaxation processes with timescales of comparable magnitudes due to coexistence of both slow and fast particles in the system. In the lane state, we observe dominant response from increased proportion of slow particles. With the increase of the field strength as the bigger lanes interact among themselves, the in-plane motion continues to slow down while they grow stronger. The heterogeneity appears when the size-distribution of the lanes is maximally broadened. Unlike the role of temperature in glasses\cite{smk-cd}, the onset of this heterogeneity is primarily due to the competition between applied field and the particle interactions\cite{lane10}. This nontrivial heterogeneous response could be verified experimentally. Also, it would be interesting to check whether these rearrangements of lanes affect the visco-elastic and dielectric response of the system, not only in the present scenario but also in cases where similar charged or magnetic dipolar colloids are subject to oscillatory field, or confinement or both \cite{appl1,appl2}, in steady states, even in ageing conditions. Moreover, we extract various time-scales of relaxations from particle tagged density relaxation. In general, the formalism revisits the work by Donati-Glotzar-Poole-Kob-Plimpton \cite{hetstr} in exploring DH a level further: It relates the heterogeneous dynamics and the heterogeneity in structural relaxation. This generic framework of exploration of heterogeneous structure is indeed novel as these structures not only relax simultaneously at different rates but also are expected to exhibit heterogeneous growth in transient conditions. Similar approach could be tested on variety of systems showing DH. On this note, we believe, this assay  opens up ranges of possibilities in unveiling unknown avenues.

\section{Acknowledgment}

The author acknowledges J. Chakrabarti for support and numerous stimulating inputs, T. Das, C. Dasgupta, S. Chatterjee and S. Sastry for discussions, S. Bose and P. Tarafdar for critically reading the manuscript. 

\section{Appendix}

\begin{figure}[h]
\includegraphics[angle=0,scale=0.15]{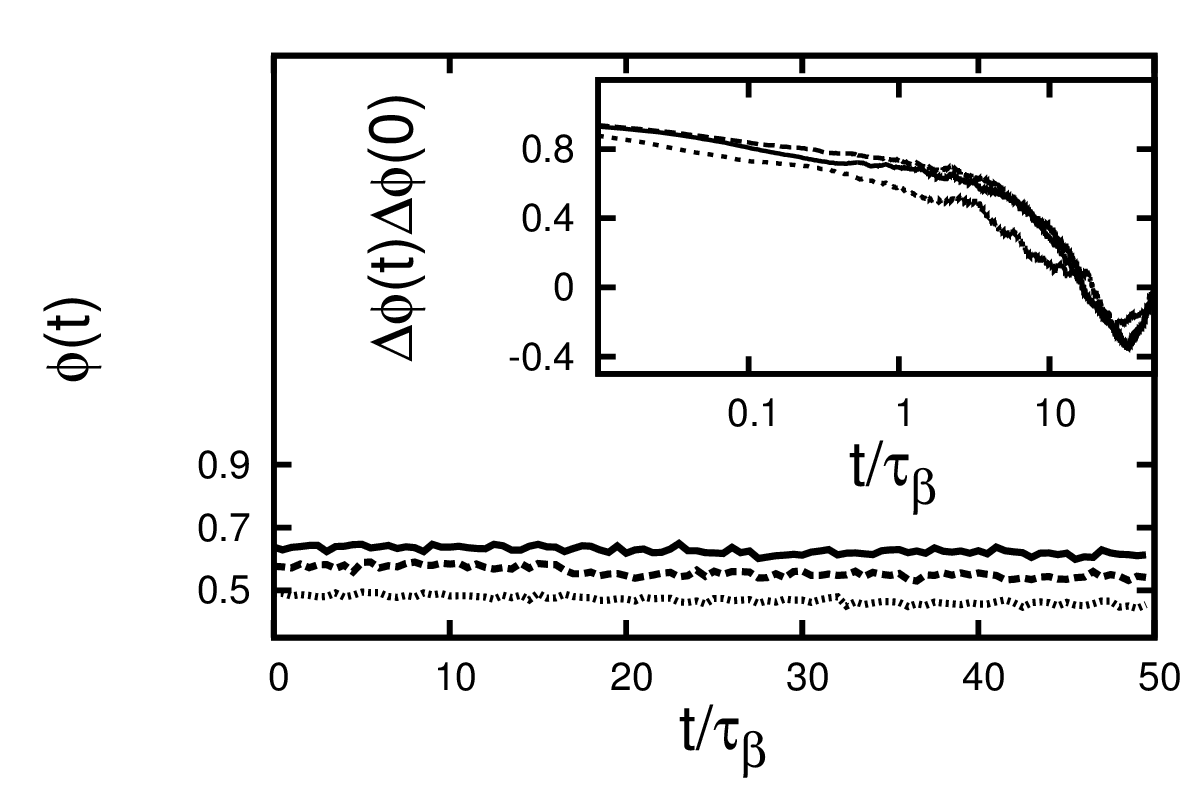}
\caption{Lane Order Parameter defined in Ref.[11], $\Phi$ in steady states for $f=50$(dotted line), $f=150$(dashed line) and $f=300$(solid line). Inset: Auto Correlation Function of $\Phi$, $\Phi(t)\Phi(0)\sim [\Phi(t)-\Phi_{S}][\Phi(0)-\Phi_{S}] $ for $f=50$(solid line),$f=150$ (dotted line)and $f=300$(dashed line)}
\end{figure}

\begin{figure}[h]
\includegraphics[angle=0,scale=0.14]{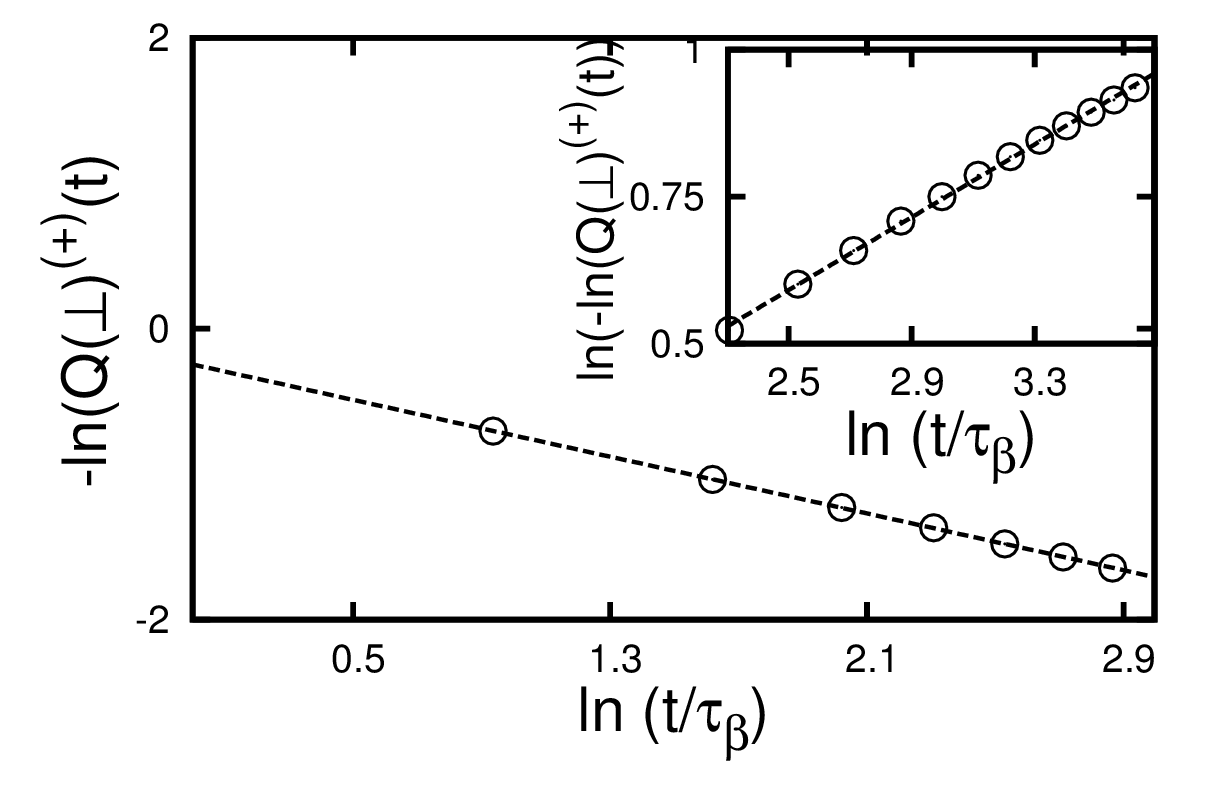}
\caption{$-ln(Q_{(\bot)}^{(+)}(t)$ vs $ln(t/\tau_{\beta})$ for $f=300$ (Main Panel) and $ln(-ln(Q_{(\bot)}^{(+)}(t))$ vs $ln(t/\tau_{\beta})$ for $f=150$ (Inset)}
\end{figure}

\begin{figure}[h]
\includegraphics[angle=0,scale=0.11]{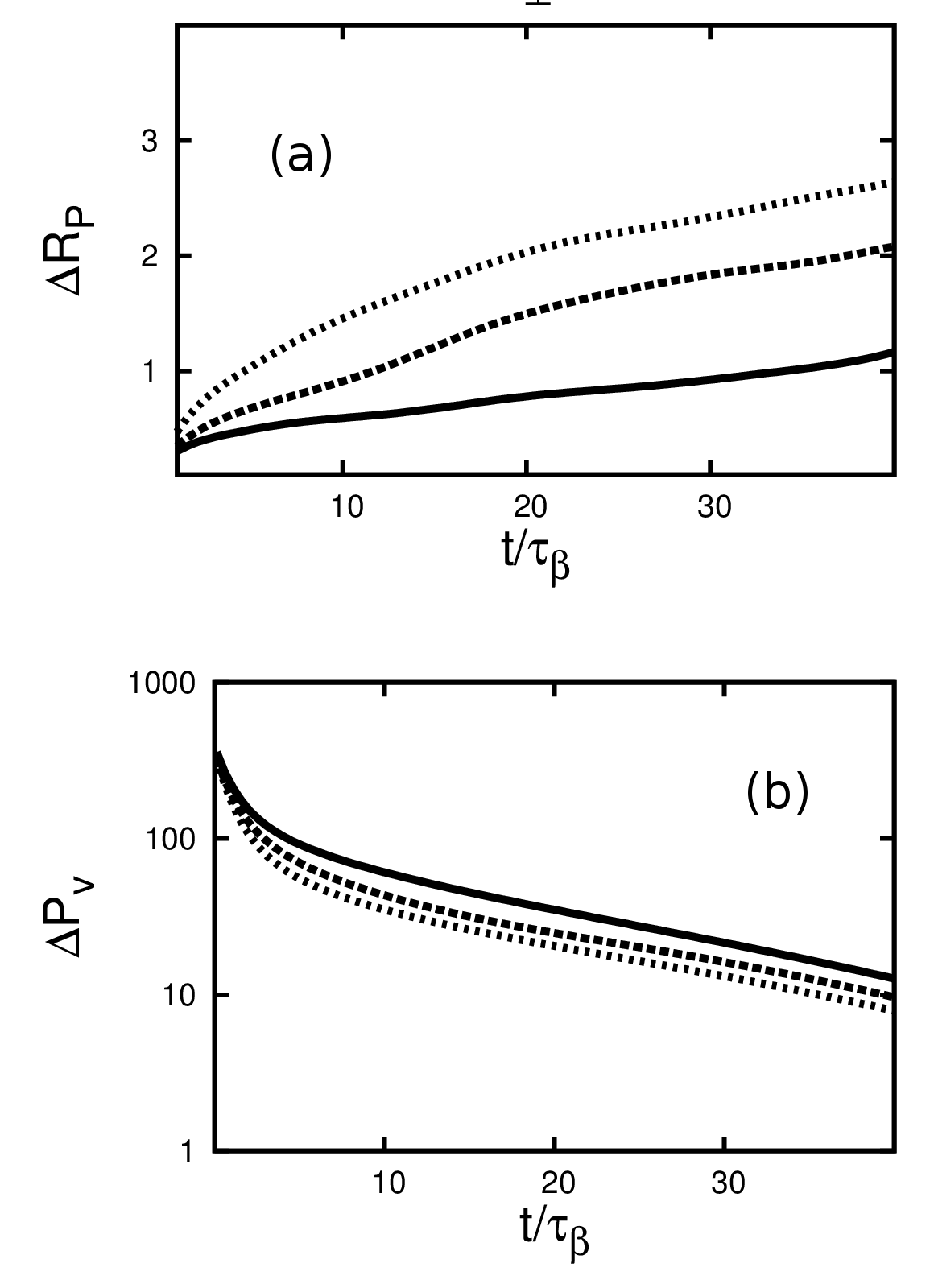}
\caption{(a)Dependence of $\Delta R_{P}$ as a function of $t$ for $f=50$(dotted line), $f=150$(dashed line) and $f=300$(solid line). (b) $\Delta P_{v}$ vs $t$ for $f=50$(dotted line), $f=150$(dashed line) and $f=300$(solid line).}
\end{figure}


\end{document}